\documentclass[twocolumn,noshowpacs,preprintnumbers,amsmath,amssymb]{revtex4-2}
\usepackage{graphicx}
\usepackage{dcolumn}
\usepackage{bm}

\begin{document}

\title{ A distributed, active patch antenna model of a Josephson oscillator }

\author{V. M. Krasnov}
\email{Vladimir.Krasnov@fysik.su.se}

\affiliation{Department of Physics, Stockholm University,
AlbaNova University Center, SE-10691 Stockholm, Sweden}


\begin{abstract}
Optimization of Josephson oscillators requires a quantitative understanding of their microwave properties. A Josephson junction has a geometry similar to a microstrip patch antenna. However, it is biased by a dc-current, distributed over the whole area of the junction. The oscillating electric field is generated internally via the ac-Josephson effect. In this work I present a distributed, active patch antenna model of a Josephson 
oscillator. It takes into account the internal Josephson electrodynamics 
and allows determination of the effective input resistance, which couples Josephson current to cavity modes in the transmission line formed by the junction. The model provides full characterization of Josephson oscillators and explains the origin of low radiative power efficiency. Finally, I discuss the design of an optimized Josephson patch oscillator, capable of reaching high efficiency and radiation power for emission into free space. 
\end{abstract}

\maketitle

\section{Introduction}

Flux-flow oscillator (FFO) is the most well studied Josephson source of high-frequency electromagnetic waves (EMW) \cite{Soerensen_1984,Nagatsuma_1985,Yoshida_1988,Winkler_1993,Golubov_1996,Koshelets_1997,Ustinov_1998,Cirillo_1998,Koshelets_2000,Krasnov_2010,Koshelets_2019,Paramonov_2022}. FFO was used in the first direct demonstration of Josephson emission by Yanson, et.al., back in 1965 \cite{Yanson_1965,Yanson_2004}. State of the art FFOs, developed by Koshelets and co-workers show a remarkable performance in terms of tunability and linewidth \cite{Koshelets_1997,Koshelets_2000,Paramonov_2022}. However, they emit very little power into free space \cite{Yanson_1965,Langenberg_1965,Langenberg_1966,Koshelets_2019}. 
The low radiation power efficiency, i.e., the ratio of radiated to dissipated power, 
is commonly attributed to a large impedance mismatch between a Josephson junction (JJ) and free space
\cite{Langenberg_1966,Krasnov_2010,MKrasnov_2021}. But there is no consensus about the value of junction impedance: is it very small \cite{Langenberg_1966}, or vice-versa very large \cite{Krasnov_2010}? At present there is no clear understanding what causes impedance mismatching and what geometrical parameters should be changed for solving the problem. Discovery of significant THz emission from stacked intrinsic JJs in layered high-$T_c$ cuprates \cite{Ozyuzer_2007,Wang_2009,Kakeya_2012,Benseman_2013,HBWang_2019,Borodianskyi_2017,Ono_2020,Delfanazari_2020,Kakeya_2020,Cattaneo_2021} further actuated the necessity of a quantitative understanding of microwave emission from Josephson oscillators. 

\begin{figure*}[th]
    \centering
    \includegraphics[width=0.95\textwidth]{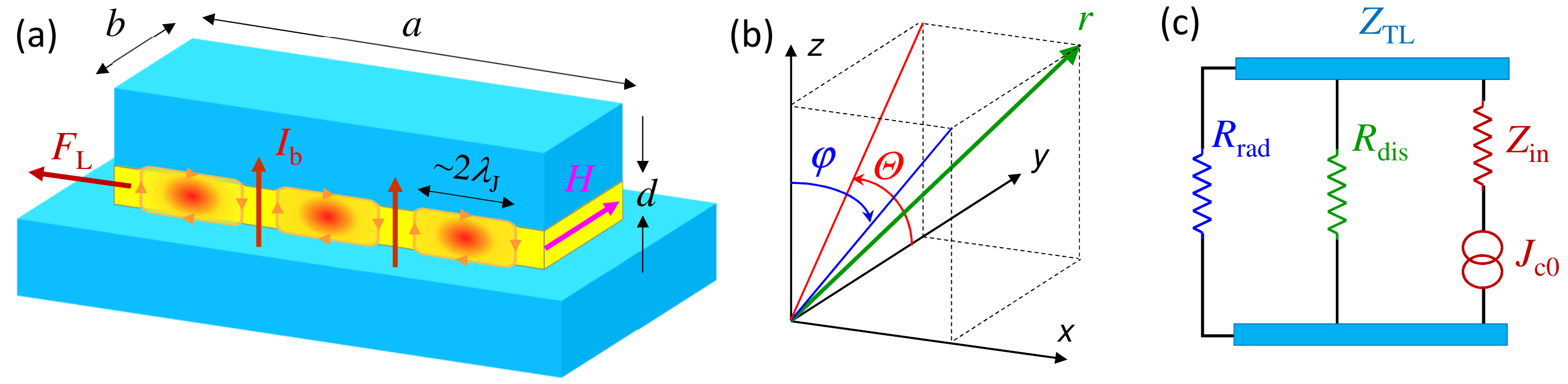}
    \caption{(Color online). (a) A sketch of the Josephson flux-flow oscillator. From outside it has a patch antenna geometry. However, inside it is driven by a distributed dc-current and the oscillating voltage is generated internally by a combination of the ac-Josephson effect and the flux-flow phenomenon. (b) Clarification of spatial and angular coordinates. (c) An equivalent circuit of the Josephson junction. The ac-Josephson effect provides a source of the high-frequency alternating current with the fixed amplitude of current density, $J_{c0}$. The oscillating voltage at the junction edges is generated by means of the input junction impedance, $Z_{\text{in}}$, and is distributed between the internal dissipative resistance, $R_{\text{dis}}$, and the external radiative resistance, $R_{\text{rad}}$, connected by the transmission line impedance $Z_{\text{TL}}$. }
    \label{fig:fig1}
\end{figure*}

Figure \ref{fig:fig1} (a) shows a sketch of a typical FFO. It is based on a sandwich-type (overlap) JJ with the length, $a\gg \lambda_J$, much larger than the Josephson penetration depth, and both in-plane sizes much larger than the thickness of the junction interface, $ d \ll b \ll a$. The in-plane magnetic field, $H_y$, 
introduces a chain of Josephson vortices (fluxons) in the JJ. 
The dc-bias current, $I_b$, exerts a Lorentz force, $F_L$, 
and causes a unidirectional fluxon motion. 
Upon collision with the junction edge, fluxons annihilate. The released energy produces an EMW pulse, which is partially emitted, but mostly reflected backwards in the JJ. Propagation and reflection of FFO pulses in the transmission line (TL), formed by the JJ, leads to formation of standing waves. The corresponding cavity mode resonances are manifested by 
Fiske steps in the current-voltage ($I$-$V$) characteristics \cite{Fiske_1965,Langenberg_1966,Kulik_1965,Kulik_1967,Barone_1982,Katterwe_2010}. FFOs exhibit sharp emission maxima at Fiske steps \cite{Yanson_1965,Koshelets_2000,Paramonov_2022}. Such a conditional emission indicates that several additional and equally important phenomena (apart from the ac-Josephson effect) are involved in FFO operation \cite{Krasnov_2010}. The excitation of high-quality factor, $Q\gg 1$, cavity modes is one of them.  

Geometry is playing a decisive role for characteristics of microwave devices. Although calculations of radiative impedances of JJs do exist \cite{Koshelev_2006}, they were not made for the FFO geometry. From the outside, the overlap JJ looks like a well known microstrip patch antenna \cite{Carver_1981,Okoshi_1985,Balanis_2005}. 
The difference, however, is inside. A standard patch antenna has a point-like feed-in port, while in a JJ the bias current is distributed over the whole area of the JJ. Furthermore, the oscillating component of the current is actively generated inside the JJ by means of the ac-Josephson effect and the flux-flow phenomenon. Therefore, a JJ can be considered as an actively pumped patch antenna with a distributed feed-in current.  

In this work I present a distributed, active patch antenna model of a Josephson oscillator. It expands the TL model of a patch antenna \cite{Balanis_2005}, taking into account the spatial distribution of the input current density in a JJ, described by the perturbed sine-Gordon equation. In the presence of magnetic field and fluxons, the oscillating current is distributed nonuniformly within the junction. This nonuniformity is essential for the FFO operation. It determines the variable input resistance, which enables the coupling of Josephson current to cavity mode resonances in the junction. The presented model allows application of many of patch antenna results and facilitates full characterization of Josephson oscillators, including the emission power, directivity and power efficiency. The model explains the origin of low power efficiency for emission in free space and clarifies what parameters can be changed to improve FFO characteristics. Finally, I discuss the design of a Josephson patch oscillator, which can reach high power for emission in free space with the optimal power efficiency, $\sim 50\%$.    

\section{The active patch antenna model} 

Spatial-temporal distribution of voltage in a JJ is described by the equation (see ch.9 in Ref. \cite{Barone_1982}):
\begin{equation}
    \frac{\partial^2 V}{\partial x^2}+ \frac{\partial^2 V}{\partial y^2} -  \frac{1}{c_0^2}\frac{\partial^2 V}{\partial t^2} = L_{\square}\frac{\partial J_z}{\partial t},
    \label{Helmholtz}
\end{equation}
where $c_0$ is the (Swihart) velocity of EMWs in the TL formed by the JJ and $L_{\square}$ is the inductance of JJ per square.  
$J_z$ is the current density through the JJ, which has Cooper pair and quasiparticle (QP) components,
\begin{equation}
    J_z=J_{c0}\sin \eta + \frac{V}{r_{\text{QP}}}.
    \label{Jz}
\end{equation}
Here $J_{c0}$ is the Josephson critical current density, $\eta$ - the Josephson phase difference and $r_{\text{QP}}=R_{\text{QP}}ab$ - the QP resistance per unit area. 

Eq. (\ref{Helmholtz}) is the 
equation for an active TL \cite{Lundquist_1971} with a distributed feed-in current density $J_z$. 
Therefore, a JJ has many similarities with the microstrip patch antenna. 
However, there are three main differences: 

(i) The feed-in geometry. A patch antenna has a point-like feed-in port, through which the oscillating current is applied \cite{Carver_1981,Okoshi_1985,Balanis_2005}. The FFO is biased by a dc current, distributed over the whole JJ area. 

(ii) The excitation scheme. A patch antenna is a linear oscillator, pumped by a harmonic signal. To the contrary, a JJ is biased by a dc-current and the oscillatory component is generated inside the JJ via the ac-Josephson effect and the flux-flow phenomenon. 

(iii) Slow propagation speed of EMWs inside the JJ, $c_0\ll c$. This is caused by a large kinetic inductance of superconducting electrodes. 
For atomic scale intrinsic JJs is layered cuprates it can be almost 1000 times slower than $c$ \cite{Katterwe_2010}. Because of that, the wavelength inside the JJ is much smaller than in free space, $\lambda\ll\lambda_0$. Therefore, a JJ corresponds to a patch antenna with extraordinary large effective permittivity, $\epsilon_r^* = (c/c_0)^2$. 

Dynamics of a JJ is described by a nonlinear perturbed sine-Gordon equation,
\begin{equation}
\frac{\partial^2\eta}{\partial \tilde{x}^2}-\frac{\partial^2\eta}{\partial \tilde{t}^2} -\alpha \frac{\partial\eta}{\partial \tilde{t}} = \sin \eta -\tilde{J_b}.   
\label{SG}
\end{equation}
It follows from Eqs. (\ref{Helmholtz}) and (\ref{Jz}), taking into account the ac-Josephson relation, $V=(\Phi_0/2\pi)\partial\eta/\partial t$. Eq. (\ref{SG}) is written in a dimensionless form with space, $\tilde{x}=x/\lambda_{\text{J}}$, normalized by $\lambda_{\text J}$,
and time, $\tilde{t}=\omega_{\text p} t$, by the Josephson plasma frequency, $\omega_{\text{p}}$, 
Here $\alpha$ is the QP damping factor, 
and $\tilde{J_b}=J_b/J_{c0}$ is the normalized bias current density, which originates from the $\partial^2V/\partial y ^2$ term in Eq.(\ref{Helmholtz}) \cite{Krasnov_2020}. In what follows, ``tilde" will indicate dimensionless variables, $\tilde{\omega}=\omega/\omega_{\text{p}}$ and $\tilde{k}=\lambda_{\text{J}} k$. Definition and interconnection between different variables is clarified in the Appendix.

\subsection{Radiative resistance of a patch antenna}

A rectangular patch antenna has two radiating slots, which correspond to the left and right edges of the JJ in Fig. \ref{fig:fig1} (a). The slots can be considered as magnetic current lines (magnetic dipoles) \cite{Derneryd_1976}. The radiation power from one slot is 
\begin{equation}
P_{1}=G_1\frac{|v(0,a)|^2}{2},
    \label{Prad1}
\end{equation}
where $|v(0,a)|$ is the amplitude of voltage oscillations at the slot ($x=0,a$) and $G_{1}$ is the radiative conductance of the single slot. Low-$T_c$ JJs are operating at sub-THz frequencies, for which the wave length in free space is large, $\lambda_0\gg b\gg d$. In this limit \cite{Derneryd_1976,Balanis_2005},
\begin{equation}
G_{1} = \frac{4\pi}{3Z_0}\left[\frac{b}{\lambda_0}\right]^2,~~~~(b\ll\lambda_0)
\label{G1}
\end{equation}
where $Z_0 =\sqrt{\mu_0/\epsilon_0} \simeq 376.73 ~(\Omega)$
is the impedance of free space. 

To calculate the total radiation power from both slots one has to take into account the mutual radiative conductance, $G_{12}$, and the array factor $AF$ \cite{Balanis_2005}. $G_{12}$ is originating from a cross product of electric and magnetic fields generated by different slots. For $\lambda_0\gg b\gg d$ it is equal to \cite{Derneryd_1978,Balanis_2005}
\begin{equation}
    G_{12}=\frac{\pi}{Z_0}\left[\frac{b}{\lambda_0}\right]^2 \int_0^{\pi}{J_0\left(k_0 a \sin\Theta \right)\sin^3\Theta d\Theta}.
    \label{G12}
\end{equation}
Here $J_0$ is the zero-order Bessel function, $k_0=2\pi/\lambda_0$ is the wave number in free space and the angle $\Theta$ is defined in Fig. \ref{fig:fig1} (b). For the $n$-th cavity mode, \begin{eqnarray}
    k_n=\frac{\pi}{a}n,~~~\omega_n=c_0 k_n, 
\end{eqnarray}
the argument of $J_0$ becomes $(c_0/c)\pi n \sin\Theta$. 
Since $c_0\ll c$, 
$k_0 a$ is small. Expanding in Eq. (\ref{G12}), $J_0(x)\simeq 1 -x^2/4$ (for $x\ll 1)$, 
we obtain:
\begin{equation}
    G_{12} \simeq G_1\left[ 1- \frac{2}{5}\left(\frac{c_0}{c}\pi n\right)^2\right], ~~~~~\left(\frac{c_0}{c}\pi n \ll 1\right).
    \label{G12b}
\end{equation}
It is seen that the mutual conductance for a JJ with thin electrodes (slow $c_0$) is not negligible and can be as big as the single slot conductance $G_1$, Eq. (\ref{G1}).  

The array factor takes into account the interference of electromagnetic fields from the two slots in the far field. It depends on the separation between the slots, $a$, the relative phase shift, $\beta$, and the direction $(\varphi,\Theta)$. 
Since radiation from a patch antenna is induced by magnetic current lines, it is more intuitive to consider the interference of magnetic fields, $H_1+H_2=AF~H_1$. For the geometry of Figs. \ref{fig:fig1} (a) and (b) it can be written as \cite{Derneryd_1978,Balanis_2005}
\begin{equation}
AF=2\cos\left[ \frac{1}{2}\left(k_0 a \sin \Theta \sin \varphi +\beta \right)\right].   
\end{equation}
Odd-number cavity modes have antisymmetric voltage oscillations, but symmetric magnetic currents, $\beta=0$. This leads to a constructive interference with the maximum $AF = 2$ perpendicular to the patch along the $z$-axis. For even modes its vice-versa, $\beta=\pi$, and a destructive interference leads to a node, $AF=0$, along the $z$-axis. 

The total emission power is
\begin{equation}
    P_{\text{rad}}=\frac{(|v(0)|^2 +|v(a)|^2)G_1 \pm 2 |v (0)| |v(a)| G_{12}}{2},
\end{equation}
where plus/minus signs are for odd/even modes, respectively. For equal amplitudes, $|v(0)|=|v(a)|$, 
\begin{equation}
    P_{\text{rad}}=\frac{|v(0)|^2}{2R_{\text{rad}}},
    \label{Prad0}
\end{equation}
with the effective radiative resistance 
\begin{equation}
    R_{\text{rad}}=\frac{1}{1\pm G_{12}/G_1}\frac{3Z_0}{8\pi}\left[\frac{\lambda_0}{b}\right]^2.
    \label{Rrad}
\end{equation}

\subsection{Determination of voltage amplitudes}

To calculate $P_{\text{rad}}$ we need voltage amplitudes at JJ edges. Within the TL model of patch antennas, $v(x)$ is obtained by decomposition into a sum of cavity eigenmodes \cite{Carver_1981}. 
For JJs a similar approach is used for the analysis of Fiske steps \cite{Langenberg_1966,Kulik_1965,Kulik_1967,Barone_1982}. 
To separate dc and ac components, we write
\begin{equation}
\eta(x,t) = kx + \omega t + \phi(x,t).
\label{fi}
\end{equation}
Here $k = 2 \pi (\Phi/\Phi_0) / a $ is the phase gradient induced by the external field, where $\Phi$ is the flux in the JJ. $\omega = 2 \pi \Phi_0 V_{\text{dc}}$ is the angular Josephson frequency proportional to the dc voltage $V_{\text{dc}}$. The last term, $\phi$, represents the oscillatory component induced by cavity modes and fluxons. This term generates the ac-voltage, which we aim to determine:
\begin{equation}
   v(x,t)=\frac{\Phi_0}{2\pi}\frac{\partial\phi}{\partial t}.
   \label{Vac}
\end{equation}

\subsubsection{Small amplitude, multimode analysis}

In the small amplitude limit, $\phi \ll 1$, a perturbation approach can be used. A linear expansion of Eq. (\ref{SG}) yields \cite{Langenberg_1966,Kulik_1965,Barone_1982}, 
\begin{equation}
\frac{\partial^2\phi}{\partial \tilde{x}^2}-\frac{\partial^2\phi}{\partial \tilde{t}^2} -\alpha \frac{\partial\phi}{\partial \tilde{t}} = \sin(kx + \omega t) + \cos(kx + \omega t)\phi -\Delta\tilde{J_b}.
\label{Perturbation}
\end{equation}
Here $\Delta\tilde{J_b} = \tilde{J_b} - \alpha \tilde{\omega}$ is the excess dc current with respect to the Ohmic QP line. It is caused by the second term in the r.h.s., which enables nonlinear rectification of the Josephson current. The excess dc current is defined as 
\begin{equation}
    \Delta I=I_{c0} \lim_{T\rightarrow \infty} \frac{1}{T}\int_0^T{dt \frac{1}{a}\int_0^a{\cos(kx + \omega t)\phi dx}}.
    \label{DI_16}
\end{equation}
The oscillatory part is described by the equation
\begin{equation}
\frac{\partial^2\phi}{\partial \tilde{x}^2}-\frac{\partial^2\phi}{\partial \tilde{t}^2} -\alpha \frac{\partial\phi}{\partial \tilde{t}} = \sin(kx + \omega t).
\label{Perturbation2}
\end{equation}
A comparison with Eq. (\ref{Helmholtz}) shows that this is the active TL equation, in which the supercurrent wave, $\sin(kx+\omega t)$, is acting as a distributed ($x,t$)-dependent drive. 

To obtain $\phi$ a decomposition into cavity eigenmodes is made \cite{Langenberg_1965,Langenberg_1966,Kulik_1965,Barone_1982,Note1}, similar to the TL analysis of patch antennas \cite{Carver_1981,Okoshi_1985,Balanis_2005}: 
\begin{equation}
\label{phi_n}
\phi(x,t)= -i e^{i\omega t} \sum_{n=1}^{\infty}{g_n \cos(k_n x)}.~~~~~~~\\
\end{equation}
Substituting it in Eq. (\ref{Perturbation2}) and taking into account orthogonality of eigenfunctions, one obtains
\begin{eqnarray}
\label{gn}
g_n=\frac{B_n + i C_n}{\tilde{\omega}^2-\tilde{k}_n^2-i\alpha\tilde{\omega}}, ~~~~~~~~~~~~~~~~~~~~~~~~~~~~~~~~~~~~~~\\ 
\label{Bn}
B_n=
\frac{\sin(k-k_n)a}{(k-k_n)a}+\frac{\sin(k+k_n)a}{(k+k_n)a},~~~~~~~~~~~~~~~~~~~~ \\ 
C_n=
-\frac{1-\cos(k-k_n)a}{(k-k_n)a}+\frac{1-\cos(k+k_n)a}{(k+k_n)a}.~~~~~~~ \label{Cn}
\end{eqnarray}
From Eq. (\ref{Vac}), voltage amplitudes at radiating slots are:
\begin{eqnarray}
    v(0)= \frac{ \Phi_0 \omega }{2\pi} e^{i\omega t} \sum_{n=1}^{\infty}{g_n},~~~~~~~~~~~~~~~~~~~~~~\\
    v(a)= \frac{ \Phi_0 \omega }{2\pi} e^{i\omega t} \sum_{n=1}^{\infty}{(-1)^n g_n}. ~~~~~~~~~~~~~~~
\end{eqnarray}

\subsubsection{Excess current}

Without geometrical resonances the dc-current, well above the field-dependent critical current, $I\gg I_c(H)$, is determined by the QP resistance, $I=V/R_{\text{QP}}$. In dimensionless units, $I/I_{c0}=\alpha V/V_{\text{p}}$, where $V_{\text{p}}=\Phi_0 \omega_{\text{p}}/2\pi$ is voltage at plasma frequency. At resonances a partial rectification of the oscillating supercurrent occurs, leading to appearance of Fiske steps in the $I$-$V$ curves. The excess dc-current, obtained from Eq. (\ref{DI_16}), is \cite{Langenberg_1966,Kulik_1965,Barone_1982}
\begin{equation}
    \Delta I = \frac{I_{c0}}{4}\sum_{n=1}^{\infty}{[B_n \text{Im}(g_n) - C_n \text{Re}(g_n)]} 
    \label{DI}
\end{equation}

\begin{figure}[t]
\begin{center}
\includegraphics[width=0.9\linewidth]{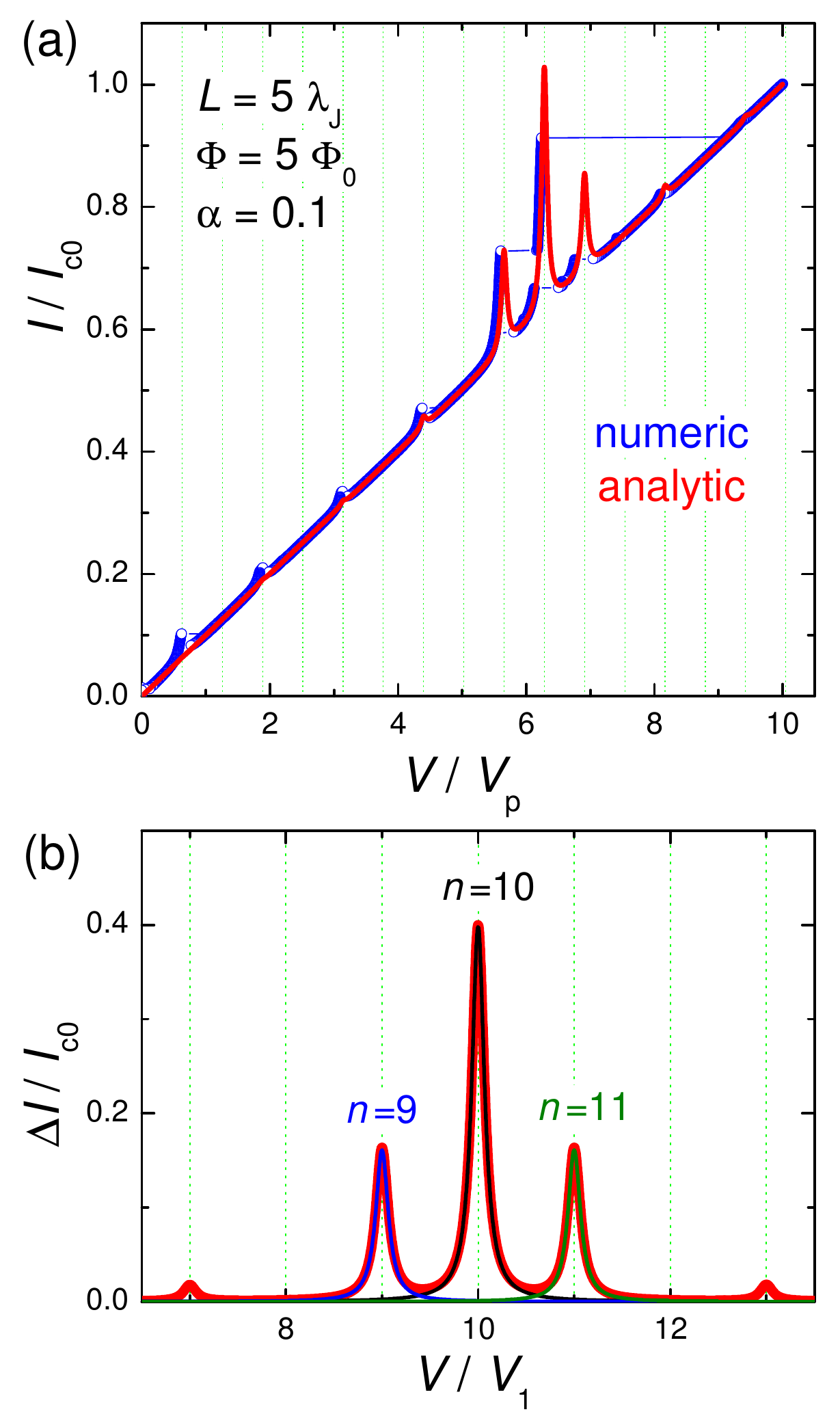}
\end{center} 
\caption{(Color online). (a) Simulated current-voltage characteristics of a junction with $L=5\lambda_J$, $\Phi/\Phi_0=5$ and $\alpha=0.1$. Blue symbols represent full numeric solution of the sine-Gordon equation (up and down current sweep). The red line represents the approximate (perturbative) analytic  solution, $I=V/R_{\text{QP}}+\Delta I$. (b) Excess dc-current, $\Delta I(V)$, at Fiske steps. Thick red line represents the multimode analytic solution, Eq. (\ref{DI}). Thin blue, black and olive lines show single mode solutions for $n=9$, 10 and 11. Vertical grid lines in (a) and (b) mark Fiske step voltages. Voltages are normalized by (a) the plasma frequency voltage, $V_p$, and (b) the lowest Fiske step voltage, $V_1$.   
}
\label{fig:fig2}
\end{figure}

Figure \ref{fig:fig2} (a) shows calculated $I$-$V$ characteristics of a JJ with $a=5\lambda_{\text J}$, $\alpha=0.1$ and at magnetic field corresponding to $\Phi=5\Phi_0$ in the JJ. Blue symbols represent direct numerical simulation of the sine-Gordon equation (\ref{SG}) for up and down current sweep. The red line shows the analytic solution, with the excess current given by Eq. (\ref{DI}). The agreement between exact (without linearization) numeric and (approximate) analytic solutions is quite good. It is seen that a series of Fiske steps appear in the $I$-$V$. Vertical grid lines mark positions of cavity mode resonances, $\omega/c_0 = k_n$. Fiske steps appear at this condition due to vanishing of $\tilde{\omega}^2-\tilde{k}_n^2$ term in the denominator of $g_n$, Eq. (\ref{gn}). The main step occurs at the double resonance condition, $\omega/c_0 = k_n = k$. It happens at $n = 2\Phi/\Phi_0$ and leads to vanishing of $(k-k_n)$ 
in the denominators of Eqs. (\ref{Bn}) and (\ref{Cn}). The condition, $\omega/c_0=k$, is referred to as the velocity matching 
because at this point the velocity of fluxon chain [or phase velocity of the current wave in Eq. (\ref{Perturbation2})] reaches $c_0$ \cite{Langenberg_1966}. 

\subsubsection{Single mode analysis}

Fig. \ref{fig:fig2} (b) shows the excess current, $\Delta I/I_{c0}$ versus $V$, normalized by the $n=1$ Fiske step voltage, $V_1= \Phi_0 c_0 / 2 a$. Such normalization clearly shows that the main resonance occurs at $n=2\Phi/\Phi_0=10$. The thick red line represents the full multimode solution, Eq. (\ref{DI}). Thin blue, black and olive lines represent a single eigenmode contribution for $n= 9$, 10 and 11. A perfect coincidence with the red line indicates that for underdamped JJs, $\alpha \ll 1$, it is sufficient to consider just a single mode. This greatly simplifies the analysis. 

For a resonance at mode $n$,  
\begin{equation}
    g_n(\tilde{\omega}=\tilde{k}_n)=\frac{iB_n - C_n}{\alpha \tilde{k}_n},
    \label{gn_1}
\end{equation}
and 
\begin{eqnarray}
 \label{v0n}
|v_n(0,a)|= \frac{\Phi_0 \omega}{2\pi}|g_n|=\frac{\Phi_0 \omega_p}{2\pi\alpha}F_n,\\
\label{DIn}
 \Delta I = \frac{F_n^2}{4 \alpha \tilde{k}_n}I_{c0},~~~~~~
 \end{eqnarray}
where 
 \begin{equation}
      F_n=\sqrt{B_n^2+C_n^2}.~~ 
      \label{Fn}
 \end{equation}

\subsubsection{Large amplitude case}

The described above perturbative approach is valid only for small amplitudes. Simulations in Fig. \ref{fig:fig2} (a) are made for an underdamped JJ, $\alpha =0.1$. In this case the quality factor of high-order cavity modes is large,
\begin{equation*}
    Q_n= \omega_n R_{\text{QP}} C = \frac{\tilde{\omega}_n}{\alpha} \gg 1,
\end{equation*}
and $|g_n|$ is not small. Since $\phi$ appears within the $\sin \eta$ term in Eq.  (\ref{SG}), the maximum possible amplitude of $|g_n|$ is $\pi$. This reflects one of the key differences between FFO and patch antenna. The patch antenna is a linear element, in which the voltage amplitude is directly proportional to the feed current. FFO is essentially nonlinear. The amplitude of Josephson phase oscillations will not grow beyond $|g_n|=\pi$. Instead higher harmonic generation will occur. 

Full numerical simulations of the sine-Gordon equation (\ref{SG}), shown by blue symbols in Fig. \ref{fig:fig2} (a), reveal that the amplitude of oscillations reach $\pi$ at the end of the velocity-matching step. This causes a premature switching out of the resonance before reaching the resonant frequency. It is somewhat miraculous that the agreement with the perturbative solution [red line in Fig. \ref{fig:fig2} (a)] is so good. Apparently, it works remarkably well, far beyond the range of its formal applicability, $|g_n| \ll 1$. 

A general single mode solution for an arbitrary amplitude was obtained by Kulik \cite{Kulik_1967}. The amplitude at the resonance, $\tilde{\omega}=\tilde{k}_n$, is given by the first solution of the implicit equation \cite{Barone_1982},
\begin{equation}
    J_0\left(\frac{|g_n|}{2}\right) = \frac{\alpha \tilde{k}_n}{F_n} |g_n|,
    \label{Kulik}
\end{equation}
where $J_0$ is the 0-order Bessel function. This equation can be easily solved numerically. It is also possible to obtain an approximate analytic solution by expanding $J_0(x)\simeq 1 -x^2/4$ for small $x$. With such expansion, Eq. (\ref{Kulik}) is reduced to a quadratic equation with the solution,
\begin{equation}
    |g_n|=\sqrt{16+\left(\frac{8\alpha \tilde{k}_n}{F_n}\right)^2}-\frac{8\alpha \tilde{k}_n}{F_n}. 
    \label{gn_2}
\end{equation}
For overdamped JJs, $\alpha \gg 1$ it reduces to the small amplitude result of Eq. (\ref{gn_1}), $|g_n|=F_n/\alpha \tilde{k}_n$. For underdamped JJs, it qualitatively correctly predicts saturation of the amplitude for $\alpha \rightarrow 0$, although at the value 4 instead of $\pi$. Thus, Eq. (\ref{gn_2}) provides a simple and  good-enough approximation for a significantly broader range of damping parameters than Eq. (\ref{gn_1}). 

\subsection{Input resistance}

For the practically most important velocity matching mode, $k_n=k$, from Eqs. (\ref{gn},\ref{Bn},\ref{Cn}) it follows, $B_n=1$, $C_n=0$, $F_n=1$, 
leading to a remarkably simple result,
\begin{equation}
    |v(0,a)|=\frac{\Phi_0 \omega_p}{2\pi\alpha} =I_{c0}R_{\text{QP}}.\\
 \label{v0k}
\end{equation}
This equation has a straightforward meaning, illustrated by the equivalent circuit in Fig. \ref{fig:fig1} (c). A JJ is a source of spatially distributed oscillating current, $J_z=J_{c0}\sin(\omega t +kx)$, with a fixed amplitude, $J_{c0}$, but spatially dependent phase, $kx$. It couples to the cavity mode via some effective input impedance $Z_{in}$. 
$Z_{in}$ depends on $\omega$, $k_n$ and $k$ and is in general complex. However, since the phase of the current wave is strongly varying along the junction, it is hard to define the phase shift between current and voltage. Therefore, in what follows I will be talking about the input resistance, $R_{in}= |Z_{in}|$, defined via the relation 
\begin{equation}
    |v(0,a)|=I_{c0} R_{\text{in}}.
    \label{Rin1}
\end{equation}

\begin{figure}[t]
\begin{center}
\includegraphics[width=0.9\linewidth]{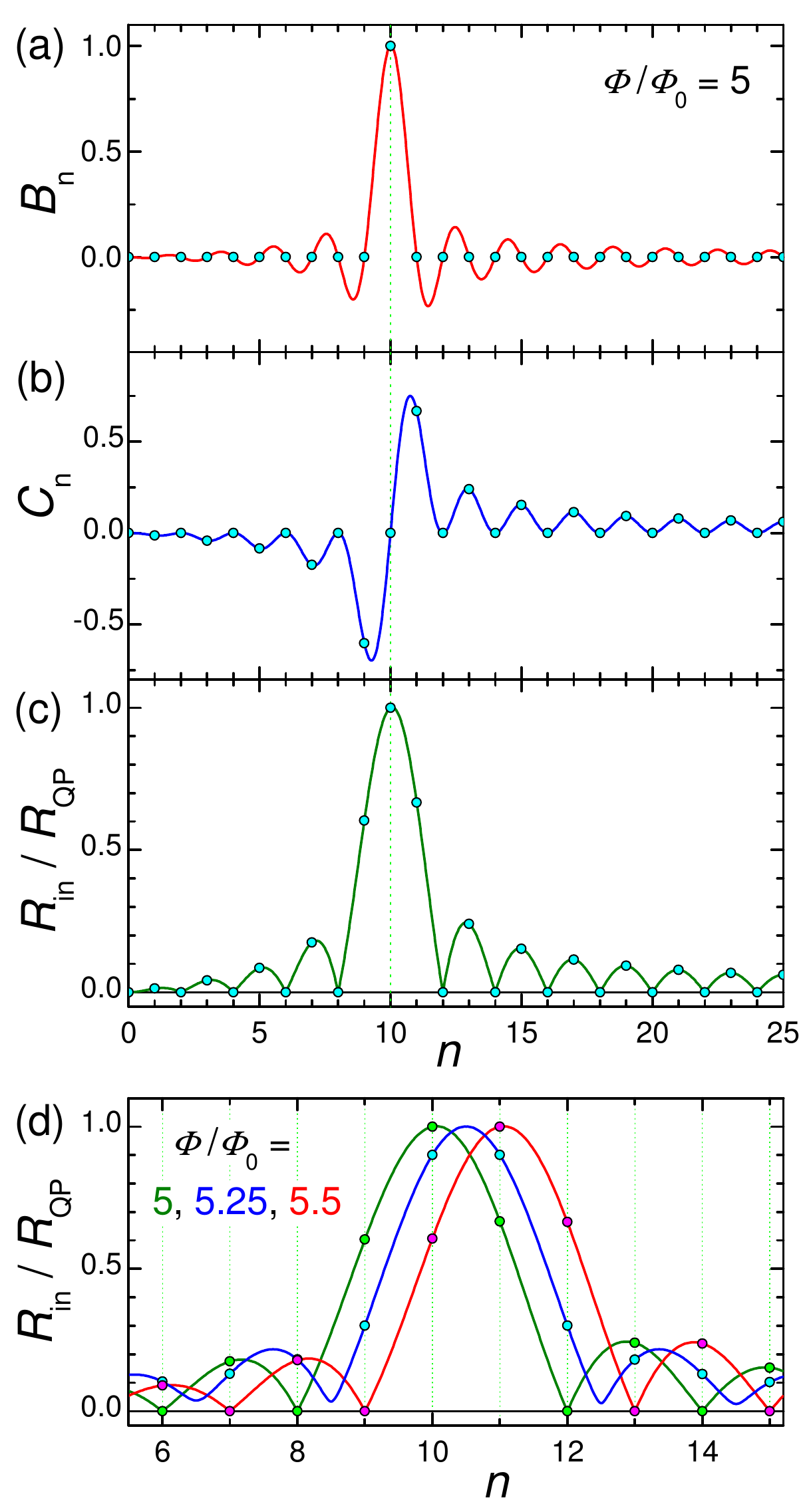}
\end{center} 
\caption{(Color online). Panels (a) and (b) show mode-number dependence of coefficients $B_n$ and $C_n$, given by Eqs. (\ref{Bn}) and (\ref{Cn}), for the case from Fig. \ref{fig:fig2} with $\Phi/\Phi_0=5$. Panel (c) shows corresponding oscillatory dependence of the input resistance, Eqs. (\ref{Fn}), (\ref{Rin2}). (d) Input resistance for $\Phi/\Phi_0=$ 5 (olive), 5.25 (blue) and 5.5 (red). The large $R_{\text{in}}$ enables good coupling of the cavity mode to the Josephson current.
}
\label{fig:fig3}
\end{figure}

From Eq. (\ref{v0n}) it follows, 
\begin{equation}
    R_{\text{in}}=R_{\text{QP}}F_n. 
    \label{Rin2}
\end{equation}

Figures \ref{fig:fig3} (a-c) show (a) $B_n$, (b) $C_n$ and (c) $R_{\text{in}}/R_{\text{QP}}=F_n$ versus $n$ for the case from Fig. \ref{fig:fig2}. Lines are obtained for continuous variation of $n$ in Eqs. (\ref{Bn},\ref{Cn}) and circles represent the actual cavity modes with integer $n$. From Fig. \ref{fig:fig3} (c) it is seen that $R_{\text{in}}$ has a distinct maximum at the velocity matching condition $n=2\Phi/\Phi_0=10$. At this point $\tilde{\omega}=\tilde{k}_n=\tilde{k}$ the wave numbers of the cavity mode and the current wave coincide, leading to a perfect coupling along the whole length of the JJ. Therefore, $R_{\text{in}}=R_{\text{QP}}$ and $v=I_{c0}R_{\text{QP}}$. For other modes, $k_n\ne k$, the coupling with Josephson current oscillations is much worse. As seen from Fig. \ref{fig:fig3} (c), it is oscillating with $n$. For the particular case with integer $\Phi/\Phi_0$, $R_{\text{in}}$ vanishes for all even modes. 
This leads to the absence of corresponding Fiske steps in Fig. \ref{fig:fig2} (a). 

The coupling of a cavity mode to the current wave in the JJ depends on magnetic field and flux in the JJ (via parameter $k$). This is illustrated in Fig. \ref{fig:fig3} (d) for $\Phi/\Phi_0=5$ [olive line, the same as in (c)], 5.25 (blue) and 5.5 (red). Although the oscillatory behavior of Fiske step amplitudes is well known \cite{Langenberg_1966,Kulik_1965,Barone_1982}, the interpretation of such behavior in terms of the input resistance makes a clear connection to the analysis of patch antennas, for which $R_{\text{in}}$ is one of the most important parameters. From this point of view, geometrical resonances with large voltage amplitudes appear only for modes coupled to the current source (Josephson oscillations) via a large input resistance, Eq. (\ref{Rin1}). As seen from Fig. \ref{fig:fig3} (d), the best coupling with maximum, $R_{\text{in}}=R_{\text{QP}}$, occurs for the velocity-matching step, $n=2\Phi/\Phi_0$. Modes with $R_{\text{in}}=0$ are not coupled to Josephson oscillations and, therefore, are not excited at all. In particular, there is no coupling to any mode in the absence of applied field, $R_{\text{in}}(H=0)=0$. That is why Fiske steps do not appear at zero field.   

\subsection{Inclusion of radiative losses in a cavity mode analysis}

Finally, in order to calculate radiative characteristics, we need to take into consideration radiative losses. 
In sec. B above only QP losses in a pure cavity eigenmode were considered. Yet, pure eigenmodes, $E_n\propto \cos (k_n x)$, $H_n\propto \sin (k_n x)$, do not emit any radiation because they do not produce  ac-magnetic fields at the edges $H_n(0,L)=0$ \cite{Balanis_2005}. Consequently, the Pointing vector is zero. In other words, eigenmodes have infinite radiative impedance, $Z_{\text{rad}}(0,L)=E(0,L)/H(0,L)=\infty$. Therefore, despite large electric fields, the radiated power $P_{\text{rad}} \propto E^2/Z_{\text{rad}}$ is zero \cite{Krasnov_2010}.  

Radiative losses can be included using the equivalent circuit, sketched in Fig. \ref{fig:fig1} (c). Voltage oscillations at the JJ edges are produced by the oscillating supercurrent via the input resistance, Eq. (\ref{Rin1}). The generated electromagnetic power is distributed between internal losses, characterized by the dissipative resistance, $R_{\text{dis}}$, and radiative losses to free space, characterised by the radiative resistance $R_{\text{rad}}$. They are connected by the transmission line impedance, 
\begin{equation}
  Z_{\text{TL}} = \sqrt{\frac{\bar{Z}_{\text{surf}}+i\omega \bar{L}}{\bar{G}_{\text{QP}}+i\omega \bar{C}}}.
  \label{Ztl}
\end{equation}
Here ${Z}_{\text{surf}}$ is the surface impedance of electrodes, $G_{\text{QP}}=1/R_{\text{QP}}$ is the quasiparticle conductance, $L$ - inductance and $C$ - capacitance of the JJ. ``Bars" indicates that the quantities are taken per unit length. For not very high frequencies and temperatures, the surface resistance of Nb electrodes is small (as will be discussed below). For tunnel JJs $G_{\text{QP}}$ is also small. In this case, 
\begin{equation}
    R_{\text{TL}} \simeq \sqrt{\frac{\bar{L}}{\bar{C}}}= Z_0 \sqrt{\frac{\Lambda d}{\epsilon_{\text r} b^2}}. 
    \label{Ztl0}
\end{equation}
It is very small because $b \gg \Lambda \gg d$ and for all practical cases can be neglected. Therefore, in Fig. \ref{fig:fig1} (c) we may consider that the dissipative and radiative resistances are connected in parallel.  
Analysis of patch antennas \cite{Balanis_2005} and numerical calculations for JJs with radiative boundary conditions \cite{Krasnov_2010} show that radiative losses can be simply included in the cavity mode analysis by introducing the total quality factor, $Q_{\text{tot}}$, of the cavity mode with parallel dissipative and radiative channels,
\begin{equation}
    \frac{1}{Q_{\text{tot}}} = \frac{1}{Q_{\text{dis}}}+\frac{1}{Q_{\text{rad}}}.
    \label{Qtot}
\end{equation}
Here $Q_{\text{dis}}$ is associated with all possible dissipative losses, such as QP resistance in the JJ as well as surface resistance in electrodes and dielectric losses, $Q_{\text{rad}}$ - with radiative losses,
\begin{equation}
     Q_{\text{dis,rad}}= \omega C R_{\text{dis,rad}}.
     \label{Qdr}
\end{equation}
Using definitions of $\alpha$ and $Q$, we can introduce a total damping factor
\begin{equation}
    \alpha_{\text{tot}}=\frac{\omega}{\omega_{\text p}}\frac{1}{Q_{\text{tot}}}=\frac{1}{\omega_{\text p} CR_{\text{tot}}},
    \label{atot}
\end{equation}
where the total resistance is 
\begin{equation}
    R_{\text{tot}}=\frac{R_{\text{dis}} R_{\text{rad}}}{R_{\text{dis}}+R_{\text{rad}}}.
    \label{Rtot}
\end{equation}

Thus, to include radiative losses, $\alpha$ and $R_{QP}$ in the equations above should be replaced by $\alpha_{\text{tot}}$ and $R_{\text{tot}}$. For the $n$-th cavity mode resonance we obtain,
\begin{equation}
 P_{\text{rad},n}=\frac{I_{c0}^2 R_{\text{tot}}^2}{2R_{\text{rad}}}F_n^2.
 \label{Prad_n}
\end{equation}
For the most important velocity matching resonance from Eq. (\ref{v0k}) we obtain
\begin{equation}
 P_{\text{rad},k}=\frac{I_{c0}^2 R_{\text{tot}}^2}{2R_{\text{rad}}},
\label{Prad_k}
\end{equation}
with $R_{\text{rad}}$ and $R_{\text{tot}}$ defined in Eqs. (\ref{Rrad}) and (\ref{Rtot}).

\subsection{Power efficiency}

The total power, dissipated in a JJ, is given by the product of dc voltage and dc current,
\begin{equation}
    P_{\text{tot}}= V I = \frac{\Phi_0 \omega}{2\pi} \left[ \alpha_{\text{dis}} \tilde{\omega} + \frac{F_n^2}{4 \alpha_{\text{dis}} \tilde{\omega}} \right] I_{c0}.
    \label{Ptot}
\end{equation}
Here the left factor is the dc-voltage and the right is the total dc-current. It contains the QP current (first term) and the rectified excess current, $\Delta I$, (second term). The latter is written using Eq. (\ref{DIn}) at the resonance condition $\tilde{\omega}=\tilde{k}_n$. It is important to note, that the nonlinear rectification occurs only inside the JJ. Therefore, the damping parameter $\alpha_{\text{dis}}$ within the JJ is used for both terms. The first term in Eq. (\ref{Ptot}) describes dissipative dc-losses, which generate only heat, $P_{\text{heat}}=V^2/2R_{\text{dis}}$. The second term in Eq. (\ref{Ptot}) describes the total power consumed by the cavity mode, $P_{\text{cav}}=V \Delta I$. Only this term is participating in radiation. From Eqs. (\ref{Rtot},\ref{Prad_n}) we obtain a well-known connection between the radiated power and the power consumed solely by the cavity mode, 
\begin{equation}
    \frac{P_{\text{rad}}}{P_{\text{cav}}}=\frac{2R_{\text{dis}}R_{\text{rad}}}{(R_{\text{dis}}+R_{\text{rad}})^2}.
\end{equation}
As usual, the maximum emission power is achieved at the matching condition $R_{\text{rad}}= R_{\text{dis}}$. 
In this case exactly one half of the cavity mode power is emitted and another half is dissipated. This is typical for antennas \cite{Balanis_2005} and is consistent with direct simulations for JJs with radiative boundary conditions \cite{Krasnov_2010}. 
Yet, the overall power efficiency is reduced by the ``leakage" QP current in Eq. (\ref{Ptot}), which just produces heat. For the $I$-$V$ curves in Fig. \ref{fig:fig2} (a), the Ohmic QP current is more than twice $\Delta I$ at the velocity matching step. Therefore, the total power efficiency, $P_{\text{rad}}/P_{\text{tot}}$, for such moderately underdamped JJ will not exceed $50/3 \simeq 17 \%$. Since the leakage current decreases with increasing $R_{\text{QP}}$, strongly underdamped JJs are necessary for reaching $\sim 50\%$ power efficiency. This is the case for Nb tunnel JJs \cite{Koshelets_2000} and for high-quality intrinsic JJs in Bi-2212 high-$T_c$ cuprates, for which the quality factor may exceed several hundreds \cite{Katterwe_2010} and $\Delta I$ can be several times larger than the leakage QP current \cite{Koshelets_2000,Katterwe_2010}. 
 
\section{Discussion}

\subsection{Estimation of parameters}

Lets estimate characteristic impedances for the case of Nb/AlOx/Nb tunnel JJs, which are used in the state of the art FFOs \cite{Koshelets_2000,Koshelets_2019}. I assume that $a=100~\mu$m, $b=10~\mu$m, $d=2$ nm, $\epsilon_r=10$, $d_1=d_2=100$ nm, the zero-temperature London penetration depth $\lambda_{L0}=100$ nm, $J_{c0}=5 \cdot 10^3$ (A/cm$^2$), $I_{c0}=J_{c0}ab=50$ mA, and the characteristic voltage $I_{c0}R_n =1$ mV. This yields, $R_n=20$ m$\Omega$, $C=44.25$ pF, $\Lambda=272.6$ nm, inductance $L^*=\mu_0\Lambda a/b = 3.43$ pH, $c_0/c =2.71\cdot 10^{-2}$. 

\subsubsection{Surface resistance}
Within the two-fluid model, surface resistance of two superconducting electrodes 
can be written as \cite{Schmidt}: 
\begin{equation}
R_{\text{surf}} \simeq \frac{a}{b}\mu_0^2\omega^2\lambda_{L0}^3\sigma_n \frac{(T/T_c)^4}{(1-(T/T_c)^4)^{3/2}}.
\label{Rsurf}
\end{equation}
Here $\sigma_n$ is the normal state conductivity. This approximation is valid for not very high temperatures, $T/T_c<0.8$. Taking typical parameters for sputtered Nb films, $\sigma_n\simeq 1.75 \cdot 10^5~ (\Omega \text{cm})^{-1}$ \cite{Krasnov_1992}, frequency $f=400$ GHz and $T/T_c=0.5$, we obtain: 
$R_{\text{surf}}\simeq 0.12~\Omega$. 

\subsubsection{Transmission line impedance}

TL impedance is given by Eq. (\ref{Ztl})
where $G_{\text{QP}}=1/R_{\text{QP}}$. 
For tunnel JJs  $R_{\text{QP}} \gg R_n$ at sub-gap voltages. I'll assume $R_{\text{QP}} = 25 R_n$, typical for Nb tunnel JJs \cite{Koshelets_2000,Koshelets_2019}. This gives, $R_{\text{QP}}=0.5~\Omega$ and $G_{\text{QP}}=2~\Omega^{-1}$. At $f=400$ GHz, $\omega L^*=8.61~\Omega$, $\omega C = 111.2~\Omega^{-1}$ and 
$Z_{\text{TL}}\simeq 0.278 +i0.0015~\Omega$. 
It practically coincides with the resistance of ideal TL, Eq. (\ref{Ztl0}). 
The value of $Z_{\text{TL}}$ is only slightly affected by ill-defined QP resistance and remains practically the same even if we use the upper limit, $G_{\text{QP}}=1/R_n$. Importantly, $Z_{\text{TL}}$ is small because of very small $d$. 

\subsubsection{Dissipative resistance}

The effective dissipative resistance is affected by all sources of dissipation, including QP and dielectric losses in the junction barrier and surface resistance in electrodes. According to Eq. (\ref{Qdr}), $R_{\text{dis}}$ is defined via the effective quality factor, $Q_{\text{dis}}$, which can be written as: 
\begin{equation}
    \frac{1}{Q_{\text{dis}}}= \frac{1}{Q_{\text{QP}}}+ \frac{1}{Q_{\text{surf}}}+ \frac{1}{Q_{\text{diel}}},
    \label{Qdis}
\end{equation}
where $Q_{\text{QP}}$, $Q_{\text{surf}}$ and $Q_{\text{diel}}$ are determined by QP, surface and dielectric losses, respectively. QP and surface resistance contribution can be accounted for using the TL analysis. The quality factor of TL is determined by the relation 
\begin{equation*}
    Q_{\text{TL}}= k_1/2k_2,
\end{equation*}
where $k_1$ and $k_2$ are real and imaginary parts of the wave number in the TL, $k=k_1-ik_2$. They are obtained from the TL dispersion relation,
\begin{equation*}
    k^2=-(R_{\text{surf}}+i\omega L^*)(G_{\text{QP}}+i\omega C).
\end{equation*}
Taking into account that $G_{\text{QP}}=1/R_{\text{QP}} \ll \omega C$ and $R_s\ll \omega L^*$, and $Q_{\text{TL}}^{-1}=Q_{\text{QP}}^{-1}+Q_{\text{surf}}^{-1}$, we obtain
\begin{eqnarray}
     Q_{\text{QP}}=\omega R_{\text{QP}}C \simeq 55.6,\\
     \label{Qqp}
  Q_{\text{surf}}=\frac{\omega L^*}{R_{\text{surf}}} \simeq 71.7
  \label{Qsurf}
\end{eqnarray}

Dielectric losses in AlOx barrier of a JJ were estimated in Ref. \cite{Gunnarsson_2013}. At $f\simeq 10$ GHz, $Q_{\text{diel}} \sim 10^4$. Although, it should reduce at $f=400$ GHz, we anticipate that it is still in the range of $\sim 10^3$. Therefore, dielectric losses are negligible, compared to QP and surface loses. Assuming $Q_{\text{diel}} = 500$ we obtain from Eqs. (\ref{Qdis}, \ref{Qqp}, \ref{Qsurf}),  $Q_{\text{dis}}= 29.48$  and $R_{\text{dis}}\simeq 0.265 ~\Omega$. It is close to the effective dissipative resistance of the TL, 
\begin{equation}
    R_{\text{dis}}\simeq \frac{Q_{\text{TL}}}{\omega C} = \frac{R_{\text{QP}}}{1+R_{\text{QP}}R_{\text{surf}}C/L^*}
\end{equation}

\subsubsection{Radiative and total resistances}

From Eqs. (\ref{Rrad}) and (\ref{G12b}), taking into account the smallness of $c_0/c$, we can write,
\begin{equation}
    R_{\text{rad}}\simeq \frac{3Z_0}{16\pi}\left[\frac{\lambda_0}{b}\right]^2.
    \label{RradApprox}
\end{equation}
Substituting $\lambda_0=750~\mu$m for $f=400$ GHz, we obtain a very large value, $R_{\text{rad}}\simeq 126.5$ k$\Omega$. Since $R_{\text{rad}}\gg R_{\text{dis}}$, the total resistance, Eq. (\ref{Rtot}), is $R_{\text{tot}}=0.265 ~\Omega \simeq R_{\text{dis}}$.

\begin{table}[t]
    \centering
    \begin{tabular}{|c|c|c|c|c|c|c|c|c|}
    \hline
    & & & & & & & & \\
    $R_n$ & $R_{\text{QP}}$ & $R_{\text{surf}}$ & $R_{\text{TL}}$ & $\omega L^*$ & $(\omega C)^{-1}$ & $R_{\text{dis}}$ & $R_{\text{rad}}$ &  $R_{\text{tot}}$ \\
    & & & & & & & & \\
    \hline
     & & & & & & & & \\
~0.02~ & ~0.5~ & ~0.12~ & ~0.28~ & ~8.6~ & 0.009 & 0.265 & 126.5k & 0.265 \\
 & & & & & & & & \\
    \hline
    \end{tabular}
    \caption{Estimation of characteristic resistances ( in Ohms) for a Nb/AlOx/Nb tunnel junction with sizes $a=100~\mu$m, $b=10~\mu$m, $d=2$ nm, $d_1=d_2=100$ nm, $J_{c0}=5000$ (A/cm$^2$), at $T/T_c=0.5$ and $f=400$ GHz.}
    \label{tab:Rtable}
\end{table}

Table \ref{tab:Rtable} summarizes characteristic resistances. 

\subsubsection{Radiation power}

From Eq. (\ref{Prad_k}) we get the maximum radiation power at the velocity matching condition, $P_{\text{rad},k}\simeq 0.7$ nW. It is much smaller than the total dc power at the velocity matching step, $\sim \Phi_0 f I_{c0} \simeq 40~\mu$W. The corresponding power efficiency $\sim 10^{-5}$ reflects the key problem for using FFO as a free-space oscillator. 


\subsection{Whom to blame?}

The very low radiation power efficiency of a JJ is colloquially attributed to ``impedance mismatching". However, so far there was no clear understanding of what is mismatching with what. A long-living misconception is that the mismatch is between the TL and free space impedances, $Z_{\text{TL}}\ll Z_0$ \cite{Langenberg_1966}. However, this is not the source of poor performance. To the contrary, it is beneficial to have a small TL impedance, connecting two radiating slots in a patch antenna \cite{Balanis_2005}. The small $Z_{\text{TL}}$ does not affect antenna performance and can be neglected. 

The real source of the problem becomes apparent from Eq. (\ref{Prad_k}). It is associated with more than five orders of magnitude mismatch between the total and radiative resistances, $R_{\text{tot}}\ll R_{\text{rad}}$, see Table \ref{tab:Rtable}. There are two main reasons for the mismatch: (i) The smallness of the junction width with respect to the free-space wavelength. The factor $[\lambda_0/b]^2$ in Eqs. (\ref{Rrad}) and (\ref{RradApprox}) leads to a very large $R_{\text{rad}}\gg Z_0$. (ii) The smallness of junction resistance, $R_{\text{QP}}\ll Z_0$. The huge mismatch indicates that a JJ alone does not work as a free-space oscillator.  

\subsection{What to do?}

Accurate matching between radiative and junction resistances is necessary for efficient emission into free space. Therefore, $R_{\text{QP}}$ should be increased and $R_{\text{rad}}$ decreased to a fraction of $Z_0$. 
However, this is not possible for the standard FFO geometry, as sketched in Fig. \ref{fig:fig1} (a). Indeed, increasing of $R_{\text{QP}}$ would require reduction of junction sizes, which would lead to even faster increase of $R_{\text{rad}}$. Alternatively, $R_{\text{QP}}$ can be increased 
by decreasing $J_{c0}$, 
but this will not reduce $R_{\text{rad}}$. Therefore, the impedance matching requires modification of the oscillator geometry. 

There are many ways of coupling a Josephson oscillator to free space. First, I note that biasing electrodes that are attached to the junction, significantly affect the net impedance. Since the total length of the electrodes (few mm) is larger than $\lambda_0$, the electrodes will reduce the net impedance and thus improve impedance matching with free space \cite{MKrasnov_2021}. Analysis of large JJ arrays demonstrated that long electrodes may act as a traveling wave antenna, facilitating power efficiency of several $\%$ at $f=0.1-0.2$ THz \cite{Galin_2018,Galin_2020}, which is much better than $\sim 10^{-5}$ estimated above for the bare junction without electrodes. In Ref. \cite{Koshelets_2019} a free-space oscillator based on an FFO, coupled to a double slot antenna, was demonstrated. Although the power efficiency was not specified, the detected of-chip signal up to 55 dB higher than the background noise was reported at $f=0.5$ THz. In Ref. \cite{Cattaneo_2021} a mesa structure, containing several hundreds of stacked Bi$_2$Sr$_2$CaCu$_2$O$_{8+\delta}$ intrinsic JJs was implemented in a turnstile antenna. A radiation power efficiency up to $12\%$ at $f\simeq 4$ THz was reported. The record high efficiency was attributed to a good impedance matching with free space \cite{MKrasnov_2021}. In Ref. \cite{Ono_2020} a  Bi$_2$Sr$_2$CaCu$_2$O$_{8+\delta}$ mesa was implemented into a patch antenna and the far-field emission at $f=1.5$ THz was reported. 

Common for all mentioned approaches is that junctions, which are small compared to $\lambda_0$ and, according to Eq. (\ref{RradApprox}), have poor coupling to free-space, are coupled to large passive elements, comparable with $\lambda_0$. These elements act as microwave antennas, enabling good impedance matching and enhancing power efficiency for emission in free space. The target parameters for such oscillator are: $f\sim 1-10$ THz, the high power-efficiency $\sim 50\%$ and high-enough of-cryostat power $> 1$ mW.         

\subsubsection{Josephson Patch Oscillator}

\begin{figure}[t]
\begin{center}
\includegraphics[width=0.99\linewidth]{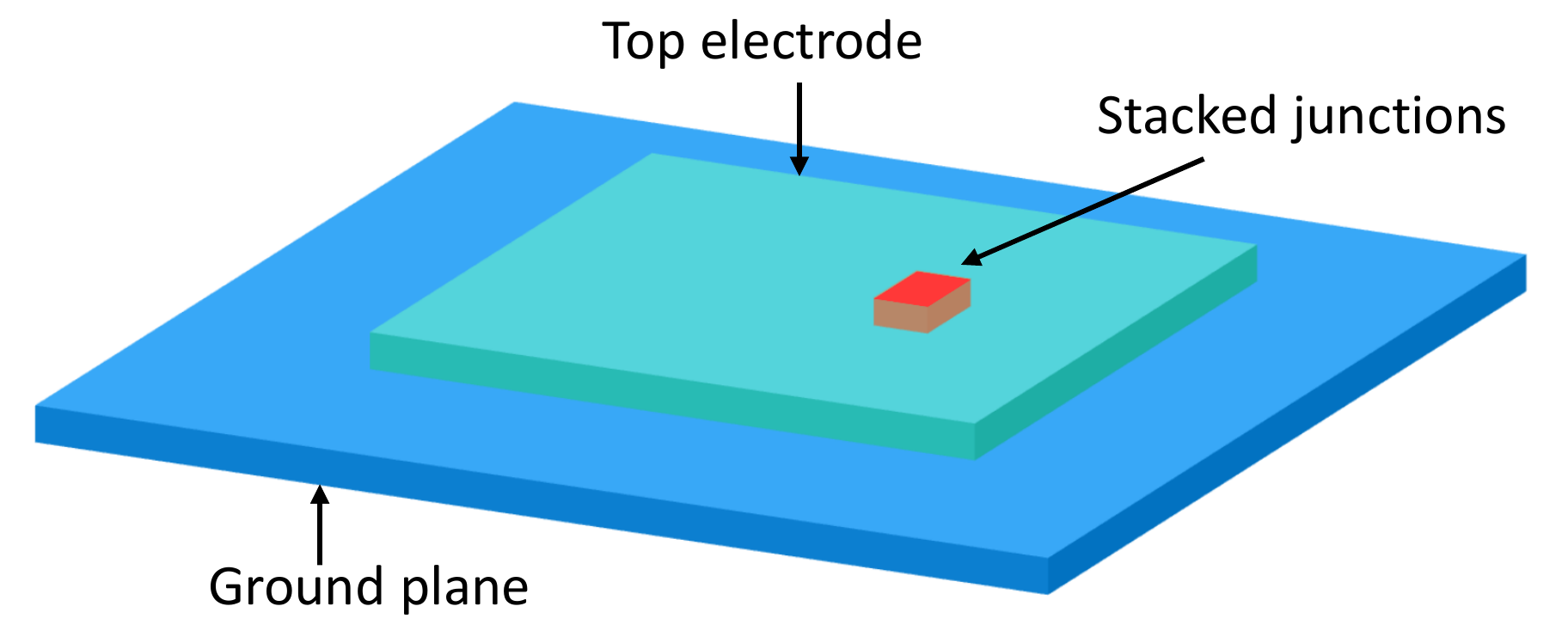}
\end{center} 
\caption{(Color online). A proposed design of the impedance-matched free-space Josephson oscillator. Here a stack of Josephson junctions is acting as source for excitation of the patch antenna formed by two large superconducting electrodes. }
\label{fig:fig4}
\end{figure}

Since in this work I consider patch antennas, below I will dwell on the patch antenna approach, discussed by Ono and co-workers \cite{Ono_2020}. Figure \ref{fig:fig4} shows a design of a Josephson patch oscillator (JPO). 
Here small junctions (red) are acting as an excitation source for a superconducting patch antenna. The bottom junction electrode (blue) forms the ground plane, and the top electrode (cyan) creates the patch antenna with sizes ($a,b$), comparable to $\lambda_0$. In principle, the JPO can be diven by a single JJ. However, as follows from the estimation above (see Table \ref{tab:Rtable}), raising the junction resistance to the desired $Z_0$ level would require a drastic (100 times) reduction of the junction area. 
This will also lead to a proportional reduction of $I_{c0}$ and the net
available power. Therefore, a better strategy is to use a stack of JJs with large-enough area, enabling high-enough $I_{c0}$. The number of JJs, $N$, is an additional controllable parameter, allowing fine-tuning of $R_{\text{n}}$ and $R_{\text{tot}}$. Furthermore, in-phase synchronization of $N$ JJs would provide the $N$-fold increment of the oscillating voltage $v(0,L)$, leading to a superradiant amplification of the emission power, $P_{\text{rad}} \propto N^2$ \cite{Krasnov_2010}. 

Moderate-size ($\sim 10~\mu$m) Bi$_2$Sr$_2$CaCu$_2$O$_{8+\delta}$ mesa structures are optimal for JPO. The $R_n$ of such mesas can be easily raised to several hundred Ohms, while maintaining $I_{c0}$ of few mA. This facilitates the optimal net power level $\sim I^2 R_n$ of several mW \cite{Cattaneo_2021,Ono_2020}. It is small enough for obviation of catastrophic self-heating, which is one of the major limiting factors for superconducting devices \cite{Cattaneo_2021,MKrasnov_2021}. Simultaneously it is large enough to enable $>1$ mW emission, provided the radiation power efficiency is close to optimal $\sim 50\%$.      

The operation frequency should be aligned with the Josephson frequency at the characteristics voltage, $I_{c0} R_{\text n}$, of JJs. 
For operation at the primary TM$^x_{100}$ mode, one side of the patch should be $a\simeq \lambda/2$, where $\lambda =\lambda_0/\sqrt{\epsilon_r}$ is the wavelength inside the  patch and $\epsilon_r$ is the relative dielectric permittivity of the insulation layer between patch electrodes. 
The other size, $b$, is adjustable and strongly affects the patch antenna performance. For $b\ll\lambda_0$ the radiative conductance per slot is given by Eq. (\ref{G1}). In the opposite limit, it becomes \cite{Balanis_2005}
\begin{equation}
    G_1=\frac{\pi}{Z_0}\left(\frac{b}{\lambda_0}\right). ~~~~(b\gg \lambda_0)
\end{equation}

One of the most important parameters of the emitting antenna is the directivity, $D$, of the radiation pattern. A rectangular patch at the TM$^x_{100}$ mode has the main lobe directed perpendicular to the patch (in the $z$-axis direction) with 
\cite{Balanis_2005}  
\begin{eqnarray*}
     D = 6.6, ~~~~~~~~~ (b\ll \lambda_0)\\
     D = 8\left(\frac{b}{\lambda_0} \right). ~~~~(b\gg \lambda_0)
\end{eqnarray*} 
A good free-space emitter should have as large $D$ as possible. From this point of view, it is preferable to have fairly wide antennas $b \sim  \lambda_0$. 

Finally, the position $(x,y)$ of the stack plays an important role in selection of the excited cavity mode. To excite solely the TM$^x_{100}$ mode the stack should be placed at $x$ close to one of the radiating slots, i.e., $x\sim a$ and $y=b/2$. The position $x$ of the stack affects the effective input resistance of the antenna and provides another adjustable parameter for patch antenna operation. The FFO input resistance, Eq. (\ref{Rin2}), is not relevant for JPO because it describes coupling to an internal cavity mode within the JJ. In JPO Josephson current is coupled to an external cavity mode in the patch. Since the patch is much larger than the JJ, the feed-in of the JPO is not distributed (in contrast to FFO). Consequently, there is no need for magnetic field. The best coupling occurs at $H=0$, corresponding to the homogeneous distribution of the Josephson current. Generally, operation of JPO is described by the standard patch antenna theory \cite{Balanis_2005}. The only interesting physics is associated with synchronization of JJs in the stack \cite{Krasnov_2010}, which can be forced by the high quality cavity mode in the antenna \cite{Cattaneo_2022}. 


\section{Conclusions}

In conclusion, I described a distributed, active patch antenna model of a Josephson oscillator. It expands the standard transmisson line model of a patch antenna, taking into account spatial-temporal distribution of the input Josephson current density in a Josephson junction. In the presence of magnetic field and fluxons, the distribution of the oscillatory component of current is nonuniform. This nonuniformity is essential for operation of a Josephson flux-flow oscillator and determines the effective input resistance, which enables the coupling between the Josephson current and the cavity modes in the junction. The presented model allows explicit application of many patch antenna results and facilitates full characterization of the device, including the emission power, directivity and power efficiency. The model explains the low power efficiency for emission in free space. It is primarily caused by the smallness of the junction width compared to the free-space wavelength, and the corresponding mismatch between very large radiative and small junction resistances. The model clarifies what parameters can be changed to improve FFO characteristics. Finally, I discussed the design of a Josephson patch oscillator, which can reach high power for emission in free space with the optimal power efficiency, $\sim 50\%$.    




\section*{Appendix}
\begin{table*}[]
    \centering
    \begin{tabular}{|c|c|c|}
    \hline
    Variable & Definition & Properties\\
    \hline
$a,~ b$ & Junction length and width in $(x, y)$ plane & $a\gg \lambda_J$, $b\sim \lambda_J$ \\
 $\alpha$ & Quasiparticle damping factor & $\alpha=1/\omega_{\text p} R_{\text{QP}}C =1/Q_{\text{QP}}(\omega_{\text p})$ \\
    $C$ & Junction capacitance &  $C=\epsilon_0\epsilon_r ab/d$ \\ 
   $c_0$ & Swihart velocity & $c_0 = c\sqrt{d/\epsilon_r \Lambda} = a/\sqrt{L^*C}$ \\ 
    $d$, $d_{1,2}$ & Thicknesses of JJ interlayer and the two electrodes & $d\ll b\ll a$ \\
    $\Phi$ & Flux in the junction & $\Phi=H_y\Lambda^* a $ \\
  $\Phi_0$ & Flux quantum & $\Phi_0 = h/2e$ \\
     $J_{c0}$, $I_{c0}$ & Maximum critical current density and critical current & $I_{c0}=J_{c0}ab$ \\
   $k$ & Field-induced phase gradient & $k=2\pi \Phi/\Phi_0$ \\
  $k_n$ & Wave number of a cavity mode & $k_n=(\pi/a)n $ \\ 
  $L^*$, $L_{\square} $  & Inductance of JJ and inductance per square & $L^*=\mu_0\Lambda a/b$,  $L_{\square}=\mu_0 \Lambda $ \\
    $\lambda_{\text{L1,2}}$ & London penetration depths of the two JJ electrodes & - \\
    $\lambda_0$ & Wavelength in free space & - \\
     $\lambda$ & Wavelength in the patch antenna & $\lambda=\lambda_0/\sqrt{\epsilon_r}$ \\
   $\lambda_{\text J}$ & Josephson penertation depth & $\lambda_{\text J}=[\Phi_0 / 2\pi \mu_0 J_{c0} \Lambda]^{1/2} = c_0/ \omega_{\text p}$  \\
  $\Lambda$ & Characteristic length associated with JJ inductance & $\Lambda=d+\lambda_{L1}\coth(d_1/\lambda_{L1}) + \lambda_{L2}\coth(d_2/\lambda_{L2})$ \\
$\Lambda^*$ & Effective magnetic thickness of the JJ & $\Lambda^*=d+\lambda_{L1}\tanh(d_1/2\lambda_{L1}) + \lambda_{L2}\tanh(d_2/2\lambda_{L2})$ \\
  $\eta$ & Josephson phase difference & - \\   
 $\omega_{\text p}$ & Josephson plasma frequency & $\omega_{\text p}=[2\pi I_{c0}/\Phi_0 C]^{1/2}$  \\
  $\omega_\text{J}$ & Angular Josephson frequency & $\omega_\text{J}= \partial \eta/\partial t = 2 \pi V_{\text{dc}}/\Phi_0$ \\ 
   $\omega_n $ & Cavity mode angular frequency & $\omega_n = c_0 k_n$ \\
    $R_{\text{QP}}$, $(r_{\text{QP}})$ & Subgap quasiparticle resistance, (per unit area) & $r_{\text{QP}}=R_{\text{QP}}ab$ \\
 $R_{\text{dis}}$ & The net dissipative resistance & - \\
 $R_{\text{surf}}$ & Surface resistance of electrodes & - \\
 $R_{\text{n}}$ & Normal state resistance of the JJ & - \\
 $R_{\text{TL}}$ & Transmission line resistance & - \\
 $R_{\text{rad}}$ & Radiative resistance & - \\
 $R_{\text{in}}$ & Effective input resistance of the JJ & - \\
 $R_{\text{tot}}$ & The total load resistance of the JJ & - \\
    \hline
    \end{tabular}
    \caption{Definition of variables (in SI units).}
    \label{tab:my_label}
\end{table*}

%

\end{document}